\documentclass[a4paper,UKenglish]{lipics}
\usepackage{microtype}%if unwanted, comment out or use option "draft"

\title{Fail Better: What formalized math can teach us about learning 
\footnote{The author acknowledges partial support by the Marie Curie project PIRSES-GA-2012-318986, funded by EU-FP7, and by CNPq~/ Brazil.}}
\titlerunning{Fail Better} %optional, in case that the title is too long; the running title should fit into the top page column

\author{Jo{\~a}o Marcos}
\affil{UFRN, Brazil\\
  \texttt{jmarcos@dimap.ufrn.br}}
\authorrunning{J. Marcos} %mandatory. First: Use abbreviated first/middle names. Second (only in severe cases): Use first author plus 'et. al.'

\Copyright{Jo{\~a}o Marcos}%mandatory, please use full first names. LIPIcs license is "CC-BY";  http://creativecommons.org/licenses/by/3.0/

% mandatory: Please choose ACM 1998 classifications from http://www.acm.org/about/class/ccs98-html . E.g., cite as "F.1.1 Models of Computation". 
% mandatory: Please provide 1-5 keywords
\serieslogo{logo_ttl}%please provide filename (without suffix)
\volumeinfo%(easychair interface)
  {M. Antonia {Huertas}, Jo\~ao {Marcos}, Mar\'ia {Manzano}, Sophie {Pinchinat}, \\
  Fran\c{c}ois {Schwarzentruber}}% editors
  {5}% number of editors: 1, 2, ....
  {4th International Conference on Tools for Teaching Logic}% event
  {1}% volume
  {1}% issue
  {119}% starting page number
\EventShortName{TTL2015}
\begin{document}

\maketitle

\begin{abstract}
Real-life conjectures do not come with instructions saying whether they they should be proven or, instead, refuted.  Yet, as we now know, in either case the final argument produced had better be not just convincing but actually verifiable in as much detail as our need for eliminating risk might require.  
For those who do not happen to have direct access to the realm of mathematical truths, the modern field of formalized mathematics has quite a few lessons to contribute, and one might pay heed to what it has to say, for instance, about: the importance of employing proof strategies; the fine control of automation in unraveling the structure of a certain proof object; reasoning forward from the givens and backward from the goals, in developing proof scripts; knowing when and how definitions and identities apply in a helpful way, and when they do not apply; seeing proofs [and refutations] as dynamical objects, not reflected by the static derivation trees that Proof Theory wants them to be.

I believe that the great challenge for teachers and learners resides currently less on the availability of suitable generic tools than in combining them wisely in view of their preferred education paradigms and introducing them in a way that best fits their specific aims, possibly with the help of intelligent online interactive tutoring systems.  As a proof of concept, a computerized proof assistant that makes use of several successful tools already freely available on the market and that takes into account some of the above findings about teaching and learning Logic is hereby introduced.  To fully account for our informed intuitions on the subject it would seem that a little bit extra technology would still be inviting, but no major breakthrough is really needed: We are talking about tools that are already within our reach to develop, as the fruits of collaborative effort.
 \end{abstract}

\section{Ask yourself again: What is a Proof?}
\label{failS}

% Language itself has a mathematical structure.

Non-obvious arguments have interesting inner structures: \textit{givens} may be assumed to be connected to \textit{goals} by directed loop-free finite paths; each node of such a structure would be annotated with some statement; the finite collection of all edges concurring in a node~$S_m$ would itself be annotated by a \textit{justification} according to which the statements~$S_1, S_2, \ldots, S_{j_m}$ contained in their other endpoints would suffice for an appropriate inferential link to be recognized; source nodes would be labelled in such a way that some of them could be identified as temporary suppositions, and the others as assumptions that would have been employed as starting points for the corresponding argument.  
If one puts the corresponding directed acyclic graph upside-down and unravels the paths to avoid a node to be visited more than once, the sink nodes become roots of their own \textit{derivation trees}.  
While checking that such an argument has been constructed flawlessly according to a number of inference rules set in advance essentially reduces to a rather costless tree traversal task, the very task of constructing the goal-directed argument, starting from the initial given hypotheses is however a potentially costly problem that might require a lot of ingenuity to be solved.
However, real-life problems will often demand practitioners to construct precisely this kind of interwoven structures concerning a given conjecture without a hint on whether this conjecture is \textit{true} or \textit{false} --- and often the arguments supporting its truth are quite distinct from the arguments supporting its falsity.  
It should be no wonder that students often have difficulty with mastering the techniques for going about and develop their own arguments, and tutors often have a hard time teaching students how to proceed.

Many tasks related to exploring conjectures by either proving or refuting them may be fully automated, as, for instance, in cases in which complete decision procedures are known to exist.  Arguably, however, if students are to be more than button-pushers for imported technological toys that play by themselves, they have more to learn by \textit{implementing} the decision procedures than by actually \textit{using} their tools to evaluate whatever conjecture needs to be checked.  At any rate, I believe students can learn many important skills even when working with fully automatable theories, as long as they approach them in a productive way, and I believe tutors have a lot to learn from the many advancements that have been made in the fast-growing field of formalized mathematics.
The teaching principles subsequently discussed in the present paper are intended to provide evidence for such beliefs.

\section{Lessons from formalized mathematics}

Mathematics in the early 19th century was navigating a sea of uncalculated logical gaps and gluts, with few safe harbors in which to dock.  Its non-systematic use of axioms and relatively careless use of definitions, found amid a baroque prose that allowed for much ambiguity and imprecision to crop in, resulted in a disgraceful state of affairs in which the mathematical vessel appeared to be going adrift.  Disaster seemed imminent, had it not been by the masters of the trade not letting it sink.  The forerunners of the now venerable foundational schools strove to provide firm grounds for Mathematics, and Logic had important roles to play in such quest.  
`Safe' proof methods and techniques were progressively identified, and stricter standards were imposed for an argument to be considered acceptable.  In the following century the very notion of \textit{proof} became a mathematical object in itself, and one whose characterization and properties were to be deeply investigated.

The works of Gentzen and other wise forerunners of Proof Theory have not only shaped the way we now think about proofs, but have heavily influenced the ulterior design of mechanized proof-assistants.  As a matter of fact, many lessons in Logic seem to have been made clearer by implementations of such proof-assistants than they had been for several earlier generations of `pure' mathematical logicians, or early type theorists --- as an example, one could argue that the difference in the roles played by free and by bound variables becomes more striking if you think of them as sorted into quantifier rules that introduce \textit{parameters} in contrast to those that introduce \textit{unknowns}.  My contention in the present paper is that paying more attention into how novice students interact with such tools for doing computer-assisted math might help understanding other important lessons that would have remained unheard.  I~propose to discuss some such largely understated lessons in what follows.

%\vspace{-2mm}
\subparagraph*{Seeing proofs [and refutations] as dynamical objects.} 
You give your student an exam to be solved with pencil-and-paper.  She has to write a convincing argument that will lead to either proving or refuting a given conjecture.  Having been rightly taught that a mathematical proof is about adequately justifying her conclusions and not about providing an explanation of her thought processes (cf.~\cite{velleman:proveit2}), the student hands back to you a scratch of the logical structure of her argument, in the form of a derivation tree.  How did she arrive to it?  You might never know.  To the best of your knowledge, she might have received the full argument ready from the dancing hands of Shiva --- or maybe from a fellow student.

This time you are the one scratching the skeleton of a proof in the blackboard, in front of your student.  Instead of cleanly proceeding from hypotheses to conclusion, or maybe proceeding the other way around, you allow yourself some false starts, backtrack from dead ends, get stuck here and there, and fail in every possible way --- having an eye at what can be learned from such repeated failures.  You exhibit the full plethora of feelings that accompany the process of inventing/discovering a proof, including the mixture of satisfaction and relief when you manage to fill all the blanks that separate givens and goals.
Your student will be happy to follow you attentively, and then take a picture of the final result with her mobile phone.  When she arrives at home, though, she tries to retrace your steps, but does not manage to reproduce the same derivation tree for the given conjecture --- or to produce any derivation tree at all.

In both situations, what is not being captured by the student is the \textit{dynamical} aspect of the proof object.  The complete derivation tree that you find at your student's exam is wholly static, and the media in which it has been registered does not allow you to replay its construction.  The picture taken by the student of the proof that resulted from your toil is only the last frame of a movie.  It might be the climax of the movie, but, more often than not, it does not allow the interested spectator to recover details of the full story that led to it.

What's worst, in the paper or in the blackboard, an entirely correct proof might have been produced in the wrong way, and you will never get a hint of that just by looking at the final derivation tree that represents such a proof.  For instance, consider the usual derivation of the quantifier shift that allows one to conclude that $(\forall x)(\forall y)P(x,y)$ and $(\forall y)(\forall x)P(x,y)$ are equivalent formulas.  Of course, to achieve your initial goal what you should do is start by fixing a certain~$y_1$ to represent an arbitrary member of your universe of discourse and set as a new goal the proof of $(\forall x)P(x,y_1)$.  Next, you fix an arbitrary~$x_1$ and set as a new goal the proof of $P(x_1,y_1)$.  So far you have invoked two applications of a logical rule often called `$\forall$-introduction'.  Because you still need to prove $P(x_1,y_1)$ from $(\forall x)(\forall y)P(x,y)$, two applications of `$\forall$-elimination' will now do, if you choose the obvious witnesses for the instantiations.  Pick your preferred proof-assistant and write the proof starting from the given, applying $\forall$-elimination (twice) and then try applying $\forall$-introduction.  You will fail: the instances that you will produce are not `arbitrary', but only specific objects of your domain; you should not expect to generalize from them.

However essential is dynamics to a proof, the standard theoretical treatises on Proof Theory still choose to simply ignore it.  
Rather than defining proofs as sequences of collections of derivations trees each with their own purpose and history, a proof is commonly identified simply with the final derivation tree that magically gather all the former trees into a whole.
%A proof is commonly identified with a derivation tree, and not with a sequence of collections of trees that lead to the final tree.  
Most of the movie has been lost in the process, as spectators just look at the final credits in awe.  Yet all practitioners know perfectly well that something fundamental is missing.

%\vspace{-2mm}
\subparagraph*{Playing and replaying proofs.}
The last example in the previous discussion has already illustrated a common feature of proof dynamics that is taken into account by a practitioner writing a so-called \textit{proof script} for a given conjecture: 
reasoning may proceed \textit{forward} from the givens or \textit{backwards} from the goals.  For instance, given two families of sets, $\mathcal{F}$ and~$\mathcal{G}$, a student might want to prove that $x\in\bigcup\mathcal{F}$ follows from $x\in\bigcup(\mathcal{F}\cap\mathcal{G})$.  Now, reasoning backwards from the initial goal, proving $x\in\bigcup\mathcal{F}$ means going through the so-called `$\exists$-introduction' rule, where the new goal would consist in finding a witness~$y\in\mathcal{F}$ such that $x\in y$.  On the other hand, using the hypothesis $x\in\bigcup(\mathcal{F}\cap\mathcal{G})$ means going through the so-called `$\exists$-elimination' rule, for in this case the procedure, reasoning forward from the given, would consist in considering an arbitrary element~$z$ of $\mathcal{F}\cap\mathcal{G}$ such that $x\in z$.  The student will easily tie the two ends of the proof, of course, when she finds out she might choose $y=z$.

``If you can't solve a problem, then there is an easier problem you can solve: find it.'' (cf.~\cite{polya:discovery-I}).
% George Pólya, Mathematical Discovery on Understanding, Learning, and Teaching Problem Solving, Volume I
Flexible proof assistants will typically allow you to solve a problem by exploring proofs in both directions, forward and backwards, to try and find that easier problem.  Using an analogy from programming language paradigms, proof languages used by proof assistants might come in a \textit{declarative} style that ``resembles program source code more than mathematics'' or in a \textit{procedural} style in which ``one holds a dialogue with the computer'' (cf.~\cite{wiedijk:ams08}).  For the obvious reasons, research efforts have been consistently devoted to developing a \textit{mathematical vernacular} for proof assistants (cf.~\cite{deBruijn:mathvern}) that resembles the formal use of natural language found in mathematical texts.\footnote{Mathematicians can be quite conservative, in fact, and even proposals to the effect that an effort should be made to employ a more disciplined and structured use of language (cf.~\cite{lamport:21st}) have for the most part appeared to fell on deaf ears.}
% N.G. de Bruijn. The mathematical vernacular, a language for mathematics with typed sets. In P. Dybjer et al., editors, Proceedings of the Workshop on Programming Languages, Marstrand, Sweden, 1987.
% Lamport, How to write a 21st century proof  
Isn't this a whole lot of work just to end up with something that appears as if it had been done with pencil-and-paper?  Well, this is where the analogy breaks down.  Note indeed that proofs described as scripts in a proof assistant can be \textit{executed} like any program.  That allows one, in principle, to ``describe not only what the machine should do, but also why it should be doing it --- a formal proof of correctness'' (cf.~\cite{gonthier:ams08}).

%\vspace{-2mm}
\subparagraph*{More on the power of computer-inspired analogies.}
Many a cognitive scientist has found a common denominator with defenders of Artificial Intelligence in explaining the difference between the mind and the brain in terms of an analogy to the difference between software and hardware.  
Imperfect as any analogy, such an explanation will already allow for debates to run, for instance, about the meaning and the roles of consciousness and emotions in the resulting picture.
More to the point, a little experience in programming or some experience with proof assistants often help bringing forth some more specific and useful analogies.  For instance, a comparison that many students find compelling is the one that equates the discharge of assumptions and the discharge of parameters in certain natural deduction proofs to the destruction of local variables, in imperative programming, when leaving the scope of functions to which they belong, and that equates non-discharged hypotheses to global variables of a program.  For another example, the familiarity with functions with type \texttt{void}, in programming, not rarely appears to be advantageous to students striving to understand the interpretations of nullary functions or nullary predicates.

Allegedly, a good education in Computer Science turns the apprentice more prone to recognizing that many symbols for variables that appear in mathematical expressions are actually bound to certain operators (such as summations, limits or integrals), or are bound to hidden (universal) quantifiers, and so they might usefully be thought of as standing as placeholders that allow for certain kinds of syntactic reductions.  From an implementation viewpoint, the conceptual understanding of the role of \textit{binding} is in fact shared by computer scientists more often with linguists than with mathematicians.  On the same track, the insightful Curry-Howard correspondence that maps propositions to types, proofs to programs, and proof simplification strategies to program evaluation strategies tends to be better appreciated by those who have been exposed to the compilation routines of modern programming languages or proof assistants.

And here is a final note on the seductive power of analogy.  As with any search related to problem solving, one way of thinking about \textit{proof search} is as another example of disciplined activity that benefits from intelligent guidance.  Now, if one is still willing to defend that ``intellectual activity consists mainly of various kinds of search'' (cf.~\cite{turing:intmac48}), then one should be more than willing to build tools to let the students experiment with and search for their own proofs, and invest on the partnership between students and their `intelligent machinery'.

% If in the end intelligence is related to proof-search, as a only a CS pioneer could have defended as early as in the 1940s (cf.~\cite{turing:intmac48}), we should of course build tools to let the students experiment with and search for their own proofs.
%Turing 1950: intelligence as a search problem... PROOF SEARCH

% Any sufficiently advanced technology is indistinguishable from magic.
% Arthur C. Clarke, "Profiles of The Future", 1961 (Clarke's third law)

%\vspace{-2mm}
\subparagraph*{When definitions apply [and when they do not].}
There is no space here, nor is it my intention, to discuss the role of definitions in the teaching and learning of mathematics.  This has been done with competence in~\cite{vinner:defAMT91}, and elsewhere.  The value of definitions that make precise certain fundamental concepts should not be underestimated: It would not be (much of) an exaggeration, for instance, to say that the whole of Logic amounts simply to the study of a couple of capital definitions, namely, those of the concepts of \textit{satisfaction}, \textit{proof}, \textit{computation}, and \textit{valid reasoning}.
% Advanced Mathematical Thinking
% Mathematics Education Library Volume 11, 1991, pp 65-81
% The Role of Definitions in the Teaching and Learning of Mathematics
% Shlomo Vinner

For all that matters, and putting aside for a moment their cognitive role in reasoning, I~may here assume definitions to be dealt with through a simple rewriting: \textit{definiens} is allowed to replace \textit{definiendum}, or vice-versa.  Assuming the student can process such rewritings while cleverly avoiding circular reasoning, assuming that the student competently grasps the content of the definitions she is supposed to work with, and also assuming that the student can make such definitions part of her technical repertoire, it still seems to me that insufficient emphasis is put on knowing what to do when a defined term, operation or relation applies as well as when it does not apply in the theory at hand.  
For instance, if set complement ($\setminus$) is defined in the usual way, what does it mean to reason \textit{from} a hypothesis of the form $c\notin a\setminus b$? (novice students often have difficulty in recognizing in this the conditional assertion `if $c\in a$ then~$c\in b$')
For a more important example, what does it mean for a statement \textit{not} to follow from a given set of hypotheses?
Many students who claim to understand, respectively, the notions of set complement and logical consequence still have difficulty giving useful answers to the two previous questions.

Much of what has been noted as a problem about dealing with definitions extends to dealing with identities (for which careless rewriting might in fact easily turn out nonterminating).  Universal statements concerning the \textit{failure} of certain identities are seldom used in theories that focus exclusively on equational reasoning.  For the interested reader, I recall though a nice example of the possible use of inequalities, in combination with automated proof search.  Recall that a Robbins algebra is a structure containing a commutative and associative binary symbol~$+$ and a unary symbol~$^\prime$ satisfying the equation $(\forall x)(\forall y)((x^\prime+y^\prime)^\prime+(x^\prime+y)^\prime)=x$.  A computerized proof that a Robbins algebra that contains an $+$-idempotent element is Boolean was quickly found by supposing, by contradiction, the \textit{denial} of another equation which was known since 1933 to axiomatize the variety of Boolean algebras.  This result ended up representing a key step in the proof, entirely uncovered by an automated prover, that \textit{all} Robbins algebras are Boolean, settling a problem that had been open for more than six decades (cf.~\cite{mccune:robbins97}).
% W. McCune, "Solution of the Robbins Problem", JAR 19(3), 263--276 (1997)

%\vspace{-2mm}
\subparagraph*{Finding the right balance between automation and our pedagogical purposes.}
Are computerized proofs legible?  And, to better start with, are they \textit{informative}?  How much will the students learn from them?  While the first question might largely amount to a puzzle for semiotics, the last question poses a challenge for educators.
In the previous paragraph I have just mentioned the `solution to the Robbins problem'.  It should be noted that this solution, which has resisted the attacks of generations of human provers, also proved to be really hard for human checkers, for the substitutions required by the use of the equations involved terms that were long and difficult to parse.  In this sense, the first computerized proofs that were automatically found seemed too unstructured and did not contribute much in insight for their intrepid human readers.

So much for fully \textit{automated} proving.  As we mentioned before, the languages used by \textit{interactive} proof assistants aim at being flexible, and often contain `tacticals' or `methods' that allow one to fine tune the amount of automation one wants to introduce into the most bureaucratic or relatively boring steps of a proof construction.  In the case of large steps done with automation, many proof assistants will allow one to inspect the structure of the proof object that has been constructed by them, hidden from the users' eyes.  The students directly involved in the very implementation of such tacticals or automated provers will no doubt learn a lot from their service to the community.  However, novice students that abuse on proofs produced by a `blast', followed by a seeming miracle, often fail to see how such proofs relate to the problems they have when confronted with the simplest mathematical conjectures.  In that sense, it would seem to me that, for pedagogical purposes, automation had better be introduced in really small doses.

Of course, it all depends on \textit{what} we intend to teach.  Take the example of classical propositional logic, and their usual analytic tableaux or sequent systems.  Both proof systems are capable of originating simple decision procedures.  However, if our purpose involves capitalizing on the conceptual distinction between formal proofs and formal semantics, tableaux tend to dangerously tread on the dividing line.  Moreover, when done on pencil-and-paper, the steps taken in the construction of tableaux are often left unjustified, as if the derivation tree by itself would be straightforward to decipher.  Arguably, and that will surely depend on the way you teach them, much more flexible systems based on sequents would also seem to pose some difficulties related to the understanding of formal semantics.
Many will contend, in fact, that natural deduction is superior as a proof system that allows students to more easily construct a conceptual map in between the way they have been taught to formulate their mathematical reasoning using regimented fragments of natural language and their conversations with the machine.
Not by chance, several interactive proof assistants in the market do make use of some variety of natural deduction in their metalanguage.
All in all, if I may insist, assuming the chosen purpose of teaching structured mathematical thinking, the motto should be: `Less automation Is More'.

%\vspace{-2mm}
\subparagraph*{On the role of proof strategies.}
Some proof systems are so robust that proofs written in them are bound to terminate (by confirming or by refutating of a given conjecture) no matter how you proceed in developing them.  The odd worst-case scenario for usual analytic tableaux for classical propositional logic, for instance, is known to be much more costly than truth-tables, but if you persist long enough in decomposing your hypotheses, you will exhaust any tableau, no longer being able in fact to further extend its branches by the addition of new data.  The first-order case, as it should be clear, is quite different in that respect: Already in the examples we presented in the above paragraphs, interchanging the order in which the quantifier rules were applied could easily lead to failed attempts at producing a proof.  

Some other proof systems will only work well with a lot of external help.  For many decidable modal logics, for instance, some of the best proof systems available need loop-checkers to put an end to their proof-search procedures
(cf.~\cite{negri:JPL2005}).
% S. Negri, “Proof analysis in modal logic,” Journal of Philosophical Logic, vol. 34, pp. 507–544, 2005.
Some classic-like tableaux for many-valued propositional logics ---all such logics being obviously decidable--- will only be assured to produce terminating proofs if you abide to certain proof strategies (cf.~\cite{ccal:mar:09a,ccal:jmar:mvol:13a}).
Here again there would seem to be a task for proof-theorists: 
The lesson concerning the importance of employing proof strategies is in fact general enough to suggest that a proof system should \textit{always} be presented associated to certain proof heuristics --- even when the choice of such heuristics turns out not to affect the use of the system in actually designing successful derivations.

Designing a proof system that only allows its user a very small number of options for the construction of her proofs is often a poor and unrealistic alternative.  Equipping the user with variegated tools-for-thought and proof techniques seems much more advisable.  If this generous education process is carried out wisely, then much longer after the student has completely forgotten what she has once proven she might still know how to approach new (or the same old) problems, and go about solving them.  The student had better learn how to construct a mathematical argument for a direct proof, an existential one, proofs by \textit{reduction}, by contraposition or by cases, learn the essential forms of induction, learn the basic useful combinatorial principles.  And she should learn how to refute a conjecture by a counter-example along with its corresponding step by step justification.  What exactly she happens to end up proving, as long as it meets her professed goal, is often less important than the increased expertise acquired by her in employing the appropriate strategies towards accomplishing the task at hand.

\section{A case study: the {\textsc{TryLogic}}}

We here briefly assess a tool that embodies some of the above discussed teaching principles.
Its design was based on a number of preexisting tools, following the IMS interoperability standards for distributed learning\footnote{See \url{http://www.imsglobal.org/lti/index.html}.}, and may be integrated as an external tool to any Virtual Learning Environment\footnote{We happened to use Moodle (\url{https://moodle.org/}) to that purpose.}.  A major interactive theorem prover\footnote{We have used a standard formal theory for \textit{natural deduction} implemented in Coq (\url{https://coq.inria.fr/}), to which I have added primitive rules for bi-implication, and we implemented in Coq also a \textit{refutation theory} for reasoning with the notion of satisfaction in formal semantics.} is used by our tool through the ProofWeb interface\footnote{We have in fact installed our own ProofWeb (\url{http://proofweb.cs.ru.nl/}) local server, and extended it to correctly display the trees from both our full natural deduction theory and our refutation theory.}.  Finally, students are intended to learn how to use the system exclusively through the \textsc{TryLogic}\footnote{Our system is available at \url{http://lolita.dimap.ufrn.br/trylogic}.} system, an infrastructure consisting of an interactive tutorial, based on a successful online system for teaching programming languages\footnote{See \url{http://try.ocamlpro.com/}.}.  Students have common access to a number of exercises to practice their skills with natural deduction, and typically they will receive directly in their areas a number of personalized exercises concerning: randomly chosen challenges involving carefully handcrafted conjectures to be proven; computer-generated exercises involving conjectures to be refuted; a number of computer-generated challenges in which they are not told whether the corresponding conjectures should be proven, or else refuted.  The latter two groups of propositional-level exercises are produced by a tool\footnote{The \texttt{Conjectures Generator} is available at \url{http://lolita.dimap.ufrn.br/logicamente-ge/}, with open source code available at \url{http://github.com/terrematte/logic-propgenerator}.} implemented over the SAT-solver Limboole\footnote{Available at \url{http://fmv.jku.at/limboole/}.}.

In my understanding, the whole process of writing proof scripts with the help of a proof assistant helps the students see proofs as instructions to be executed in order to dynamically explore a problem-solving challenge, even if that does not perfectly match the static proof-theory that they learn from the mathematical textbooks.  The difficulty of the task, in which we strictly control the amount of automation that we allow the students to use, makes them ponder about how useful it may be to entertain a strategical approach to a given challenge, and the nature of the task and the easy interaction with and responsiveness of the underlying media encourages them to do some heuristic exploration through trial and error.

Before \textsc{TryLogic}, our group had invested for a few years on the \textsc{Logicamente} (cf.~\cite{ter:cos:mar:logicamente-TICTTL11}), a growing collection of Learning Objects combined with the respective learning scripts, expositions, tasks and activities on subjects of Logic.  The latter suite of tools was largely independent from other existing tools, and was developed \textit{ab nihilo} through the collaborative effort of students that got involved in the project and were evaluated by their contributions to it. 
My impression about the project \textsc{Logicamente}, as fun as it might have been to work on, is that it was much more successful, from a pedagogical viewpoint, for the students that \textit{implemented} its modules than for other students who \textit{used} these modules later on.
I am not trying now to sell \textsc{TryLogic} (and its associated tools) to you, even though I would be delighted if you bought it (and it is free of cost).
What I feel to be really important about this new project (which we report upon in more detail in~\cite{ter:mar:trylogic-TTL15}) is that it is entirely based on exploring the teaching principles discussed in the previous section, for whose convenience and effectiveness I do not intend to offer here independent evidence except for the fact that a few dozen students have had very fruitful interactions with \textsc{TryLogic} along the last three years.
% 2012.2 - 14 out of 22 students
% 2013.1 -  7 out of 17 students
% 2013.2 - 14 out of 24 students
% 2014.2 -  8 out of 11 students
% 2015.1 - 7 out of 11 students
% TOTAL, so far: 43 out of 74 students

\section*{Coda}
%If our purpose in higher education happens to transcend the teaching of Logic as a subject in itself, but actually encompasses providing instruction concerning the \textit{use} of Logic in formalizing and structuring arguments and in solving conjectures either by proving the corresponding statements or by properly refuting them, it should be clear that we had better let our intuitions be supplemented by data provided by the actual didactic experience of the experts on the field, and allow thus our formal theories be enhanced by elements that could have been deemed unnecessary for the purely abstract expression of their contents.

%% FOR FURTHER REFLECTION
The critical transition from high school to higher education has ---and with a good reason--- received increased attention from educators worldwide.  
Two distinct paths might be taken by in individual undergoing post-secondary education, namely: focusing on technical content that adds to \textit{know how}, or investing on a scientific education that presupposes a frame of mind not dispensing with the \textit{know why}.  It has been eloquently argued elsewhere that Logic has a secure role in teaching proof and that this would be better promoted by emphasizing general principles leading to deeper understanding, rather than by focusing on concrete narrow problem-solving strategies; more generally, teaching logical reasoning ``builds students' confidence in the rationality of the mathematical enterprise and helps allay their fear of failure'' (cf.~\cite{epp:role:AMM}).

All education starts from personal reflections.  The present paper has collected my own reflections, at the present date, on the topic of `tools for teaching logic', and what we can learn from and with such tools.  Scientific discussions concerning education, though, cannot end at the level of intuition.  We need data.

In the above I have proposed that Proof Theory should redefine proofs as dynamical objects, have argued about the advantages of treating proofs as executable code, and of testing the understanding of definitions by inquiring about the conditions in which they fail to apply, have commented upon the contributions given by certain computer-inspired analogies, and finally I have briefly discussed the importance of exercising one's mathematical thinking by acquiring a number of skills polished by the intelligent use of proof strategies.
We should always aim at achieving maximal pedagogical effect.  The best way of measuring this, however, is probably not by bragging about how well the students succeed in the tasks we assign them with our own tools and measure with our own rulers, but by their later performance in other tasks that demand similar competences.  Still, data needs to be collected to test whether our teaching principles are sound and far-reaching.

If arguments aim at convincing, the same effect may be achieved \textit{ad baculum}.  We had better not teach our students to resort to sticks, though.
Many novice students have serious trouble telling the difference betweeen a proof and, say, a broccoli plantation.  Talent and experience are needed for recognizing both successful and failed proofs.
In the same way that you should not dare teaching formal grammar to a student that is still being alphabetized, I believe that students had better be repeatedly exposed to real-world proofs before you endeavor teaching them formal logic and proof techniques.
The particular tool I have described in the previous section was introduced in blended learning, and used concepts of Cognitive Load Theory in its instructional design.  I tend to believe, however, that the teaching principles discussed in this paper can ---and should--- be carried over to corresponding theoretical investigations, as well as to other tools and learning theories.

%\subparagraph*{Acknowledgements.}

%Patrick Terrematte is the main responsible for the system \textsc{TryLogic}, and Elias Amaral implemented the \texttt{Conjectures Generator}.
%Several classes of students have gone through the experience of using our tools, and I can only thank them heartfully for their feedback and their patience.
%This research has been partially funded by CNPq, and by the Marie Curie project GeTFun (PIRSES-GA-2012-318986 funded by EU-FP7).

%%
%% Bibliography
%%

%% Either use bibtex (recommended), but commented out in this sample

%\bibliographystyle{plain}
%\bibliography{try}

%% .. or use bibitems explicitely

\end{document}